\documentclass[a4paper,11pt]{article}

\usepackage{pos}
\usepackage{lipsum} 
\usepackage{makecell}    
\usepackage{sidecap} 
\usepackage{subcaption} 

\usepackage{lineno}

\title{Neutrino flavor composition using High Energy Starting Events with IceCube}

\ShortTitle{Neutrino flavor composition using High Energy Starting Events with IceCube}

\author{The IceCube Collaboration \\{\normalsize \normalfont(a complete list of authors can be found at the end of the proceedings)}\\}

\emailAdd{thijs.van.eeden@desy.de}
\emailAdd{markus.ackermann@desy.de}

\abstract{

Astrophysical neutrinos provide crucial insights into their sources and play a key role in multi-messenger astronomy. The neutrino flavor composition at Earth allows us to probe the mechanisms of neutrino production and cosmic ray acceleration, as well as the properties of the environments in which they originate. Understanding the flavor composition also offers a unique opportunity to test new physics in the neutrino sector. The IceCube Neutrino Observatory consists of 1~km$^3$  of ice instrumented with photomultipliers that detect neutrinos through Cherenkov radiation from their interaction products. Different neutrino interactions result in distinct event topologies, such as tracks, cascades, and double cascade events, which allow for the identification of the interacting neutrino type and measurement of the flavor composition of the astrophysical neutrino flux. In this contribution we present the results of the measurement of the flavor ratio of the High-Energy Starting Event Selection based on 12 years of data, a high-purity sample of neutrino interactions that occur inside the detector. In addition, we discuss various methods that could further improve the analysis in the future.

\vspace{4mm}

{\bfseries Corresponding authors:}
Neha Lad$^{1}$, 
Thijs Juan van Eeden$^{1*}$,
Markus Ackermann$^{1}$\\
{$^{1}$ \itshape DESY}\\[4mm]
$^*$ Presenter
}

\ConferenceLogo{PoS_ICRC2025_logo}

\FullConference{39th International Cosmic Ray Conference (ICRC2025)\\
 15–24 July 2025\\
Geneva, Switzerland\\}

\begin{document}


\maketitle

\section{Introduction}



High-energy neutrinos are produced in extreme astrophysical environments when cosmic rays interact with matter or photons. The neutrino flavor composition measured at Earth relates to the flavor composition at the source, which provides insights into the production mechanisms at their sources. The neutrino flavor composition also serves as a probe for new physics, enabling constraints on phenomena such as CPT violation and non-standard neutrino interactions \cite{arguellesflavor}.

The IceCube Neutrino Observatory is a neutrino detector buried in the ice at the South Pole \cite{icecubegeneral}. It consists of 1~km$^3$of ice instrumented with optical modules containing photomultiplier tubes that detect Cherenkov radiation from neutrino interaction products. The flavor composition measurement is performed using the high-energy starting event (HESE) sample \cite{hese2013}. The HESE sample yielded the first evidence of astrophysical neutrinos and has a high neutrino purity. Furthermore, it covers the full sky and contains all neutrino flavors which allows to constrain the flavor composition.

This contribution contains the updated results on the neutrino flavor composition using 12 years of IceCube data, which includes advanced simulations and event reconstruction and an improved ice model. In addition, novel methods are presented to improve this analysis in the future. This includes improving the constraints by applying a new binning method and increasing the number of neutrino events through combining the HESE sample with other event selections. Additionally, enhancing the reconstruction and classification of double cascade events will lead to a more robust and reliable analysis.

\section{Selection and particle identification}

The HESE sample utilizes a lower bound of 6000 detected photoelectrons and a veto for light in the outside layer of the detector to reduce atmospheric backgrounds. The backgrounds consist of atmospheric muons, and conventional and atmospheric neutrinos. The cosmic ray model for all atmospheric backgrounds is H4a \cite{h4a} using the SYBILL2.3c hadronic interaction model \cite{sibyll}. The conventional neutrino flux originates from the decay of pions and kaons, while the prompt neutrino flux originates primarily from the decay of short-lived charmed mesons. 

The data and Monte Carlo simulations of neutrinos and atmospheric muons are reconstructed using maximum likelihood reconstruction methods. These methods test three separate source hypotheses:
\begin{itemize}
    \item Track: $\mu$ from $\nu_\mu$ or $\nu_\tau$ charged current interactions where the $\tau$ decays into a $\mu$ (BR=17\%),
    \item Cascade: electromagnetic cascades from $\nu_e$ charged current interactions or hadronic cascades from all-flavor $\nu$ neutral current interactions,
    \item Double cascade: $\nu_\tau$ charged current interactions where the $\tau$ lepton decays into an electron or intro hadrons (BR=83\%).
\end{itemize}
The key difference is the presence of a relativistic muon that travels substantial distances while emitting light. The electrons and hadrons that produce electromagnetic and hadronic showers deposit all their energy in a relatively small region of space compared to the detector spacing. The short lifetime of the tau ($2.9 \times 10^{-13}$ s) causes the neutrino interaction and tau decay vertex to overlap at low energies, but the expected decay length increases with the tau energy $E_\tau$ due to time dilatation and is given by $L_\tau \sim 50 \text{ m} \cdot E_\tau \text{ / PeV}$.

The reconstructed properties and three likelihoods for each hypothesis enable the classification of events into one of the three topologies, as described in Ref. \cite{usertauicrc}. In order to further reduce atmospheric backgrounds and to break the degeneracy between single and double cascades, events are only selected if the total deposited energy $E_{\rm tot} > 60$ TeV. Subsequently, the classification chain applies a series of selection criteria to identify double cascade candidates. If the criteria are not met, the events are selected as cascade or track instead. The first criteria cover the reconstruction quality and are given by:
\begin{itemize}
    \item Double cascade reconstruction converged \cite{julianahese7.5},
    \item Energy of first and second cascade $E_1, E_2 \geq 1$ TeV,
    \item Both cascades have vertices no more than 50 m outside of the detector,
    \item Angle between reconstructed direction of double cascade and track reconstruction $\leq 30^{\circ}$.
\end{itemize}
If any of these criteria are not satisfied, the event is classified as a track or cascade based on the larger of the two likelihoods. The final steps are defined below.
\begin{itemize}
    \item The reconstructed tau length $L_{\rm reco} \geq 10\,\mathrm{m}$. If this condition is not met, the event is classified as a single cascade.
    \item The energy confinement, defined as $E_{\rm C} = \frac{E_1 + E_2}{E_{\rm tot}} > 0.99$. Events that do not meet this criterion are classified as tracks. The reconstructed cascade energies $E_1,E_2$ are obtained using the double cascade reconstruction, while $E_{\rm tot}$ using a dedicated algorithm that determines all energy depositions using the best-fit hypothesis. 
    \item The energy asymmetry $-0.98 \leq E_{\rm A} = \frac{E_1 - E_2}{E_1 + E_2} \leq 0.3$. Events not meeting this condition are also classified as single cascades.
\end{itemize}
The selection criteria were derived by analyzing the simulated signal and background distributions of these parameters. The resulting events from simulations and data are used in the analysis.


\section{Method}

The flavor composition is measured using a forward-folding likelihood fit. The likelihood fit is performed for the three topologies simultaneously using two-dimensional probability density functions obtained from the simulations. The observables for the double cascades are represented in logarithmic bins of the total reconstructed energy and the reconstructed tau decay length. The energy spans from 60 TeV to 12.6 PeV using 13 bins and the tau length from 10 to 1000 m using 10 bins. The observables for tracks and single cascades are the reconstructed energy and the reconstructed zenith angle. The energy spans the same energy as for the double cascades using 21 logarithmic bins, and the zenith angle is represented by 10 bins in cosine space from -1 to 1. 

The likelihood is defined in Ref. \cite{saylikelihood}, which takes limited Monte Carlo statistics into account. The total likelihood for all topologies is obtained by multiplying the separate likelihoods according to $\mathcal{L} (n |\theta, \xi) = \mathcal{L}^{\rm Double} \mathcal{L}^{\rm Single} \mathcal{L}^{\rm Track}$, where $n$ denotes the number of observed events, $\theta$ represents the signal parameters, and $\xi$ the nuisance parameters. The best-fit parameters are obtained by varying the signal and nuisance parameters. The signal parameters include the normalization of the all-flavor neutrino flux and the spectral index of a single power law: $\Phi_{\nu+\bar{\nu}} \big(\frac{E_\nu}{\text{100 TeV}}\big)^{-\gamma}$. This is complemented by two flavor fractions $f_\alpha$, where $f_{\nu_e}+f_{\nu_\mu}+f_{\nu_\tau}=1$. The atmospheric flux nuisance parameters are described in Ref. \cite{abbasi2022improved}.

Detector systematics are taken into account using the SnowStorm method \cite{snowstorm}, where every systematic is varied for each event during simulations. This allows for reweighting of the simulations depending on detector systematics. The detector systematics incorporated in the fit follow those described in Ref. \cite{abbasi2022improved}, with the addition of a scaling factor that accounts for the anisotropy in light scattering within the ice as described in Ref. \cite{julianahese7.5}.

\section{Results}

The HESE sample for 12 years of data contains 97 events with energies higher than 60 TeV. The classification of the events is shown in Table \ref{tab:pid_comparisons}.

\begin{table}[h!]    
    \centering
    \begin{tabular}{ c|c c c c|c }
        \hline
        \makecell{Reconstructed \\ Morphology} & \multicolumn{4}{c|}{$f_{\nu_e}:f_{\nu_{\mu}}:f_{\nu_{\tau}} = 0.19:0.43:0.38$} & Data \\
        & Astro & Conv & Muon & Total & \\
        \hline\hline
        Cascades & $57\pm2$ & $6\pm0.79$ & -- & \textbf{$63.4\pm2.4$} & \textbf{64} \\
        Tracks & $16\pm0.8$ & $5.7\pm0.9$ & $1.84\pm2.7$ & \textbf{$23.4\pm3.4$} & \textbf{28} \\
        \makecell{Double \\ Cascades} & $3.8\pm0.3$ & $0.3\pm0.2$ & -- & \textbf{$4.1\pm0.4$} & \textbf{5} \\
        \hline\hline
    \end{tabular}
    \caption{The expected and observed number of HESE events, classified into three detection channels, assuming the best-fit flavor composition. The total expected event counts are broken down into contributions from Astrophysical (Astro), Conventional atmospheric neutrinos (Conv), and Atmospheric Muons (Muon). The prompt atmospheric neutrino component is omitted, as the best-fit normalization ($\Phi_{\mathrm{prompt}}$) is 0.}
    \label{tab:pid_comparisons}
\end{table}

\begin{SCfigure}[1.2][h!]
  \centering
  \includegraphics[width=0.65\linewidth]{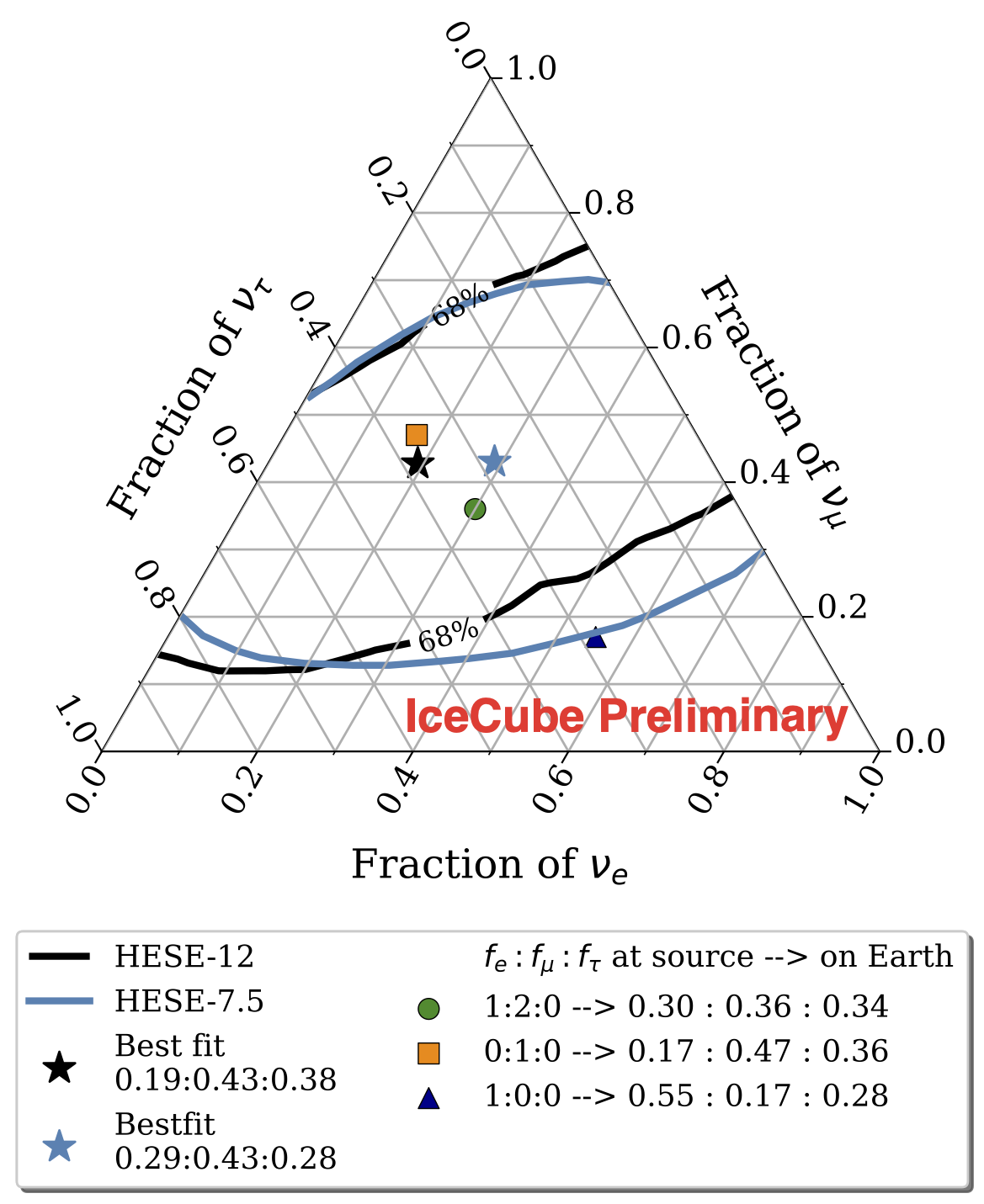}
  \caption{The best-fit flavor composition using 12 years of HESE data compared with the results using the same likelihood and 7.5 years of HESE data \cite{julianahese7.5}. The solid and lines represent the 68\% confidence regions obtained using Wilk's theorem. Expected flavor compositions on Earth from different astrophysical sources are also marked.}
  \label{fig:flavortriangle}
\end{SCfigure}

The result and the contour of the flavor composition measurement using 12 years  of HESE data is shown in Figure \ref{fig:flavortriangle}. Several neutrino flavor composition scenarios are also shown, indicating both the flavor ratios at the source and the corresponding compositions at Earth \cite{arguellesflavor}. The best-fit flavor composition is $f_{\nu_e}:f_{\nu_\mu}:f_{\nu_\tau} = 0.19^{+0.26}_{-0.15}:0.43^{+0.27}_{-0.17}:0.38^{+0.37}_{-0.24}$. The spectral index is $2.84^{+0.19}_{-0.18}$ and is consistent with the previous HESE sample results using 7.5 years of data \cite{hese7.5, julianahese7.5}. To ensure a fair comparison, the constraints for HESE-7.5 are shown using a similar likelihood, rather than the extended likelihood that includes the information from resimulating the tau neutrino candidates.

The normalization of the all-flavor neutrino flux is $\Phi_{\nu+\bar{\nu}} = 5.94^{+5.64}_{-4.28} \cdot 10^{-18} \text{ GeV}^{-1} \text{ s}^{-1} \text{ sr}^{-1} \text{ cm}^{-2}$. The larger uncertainty compared to the HESE-7.5 results in Ref. \cite{hese7.5} is expected because this analysis includes using extra free parameters for constraining the flavor composition. The expected number of events assuming the best-fit model parameters is shown in Table \ref{tab:pid_comparisons}.

\begin{SCfigure}[1.2][h!]
  \centering
  \includegraphics[width=0.72\linewidth]{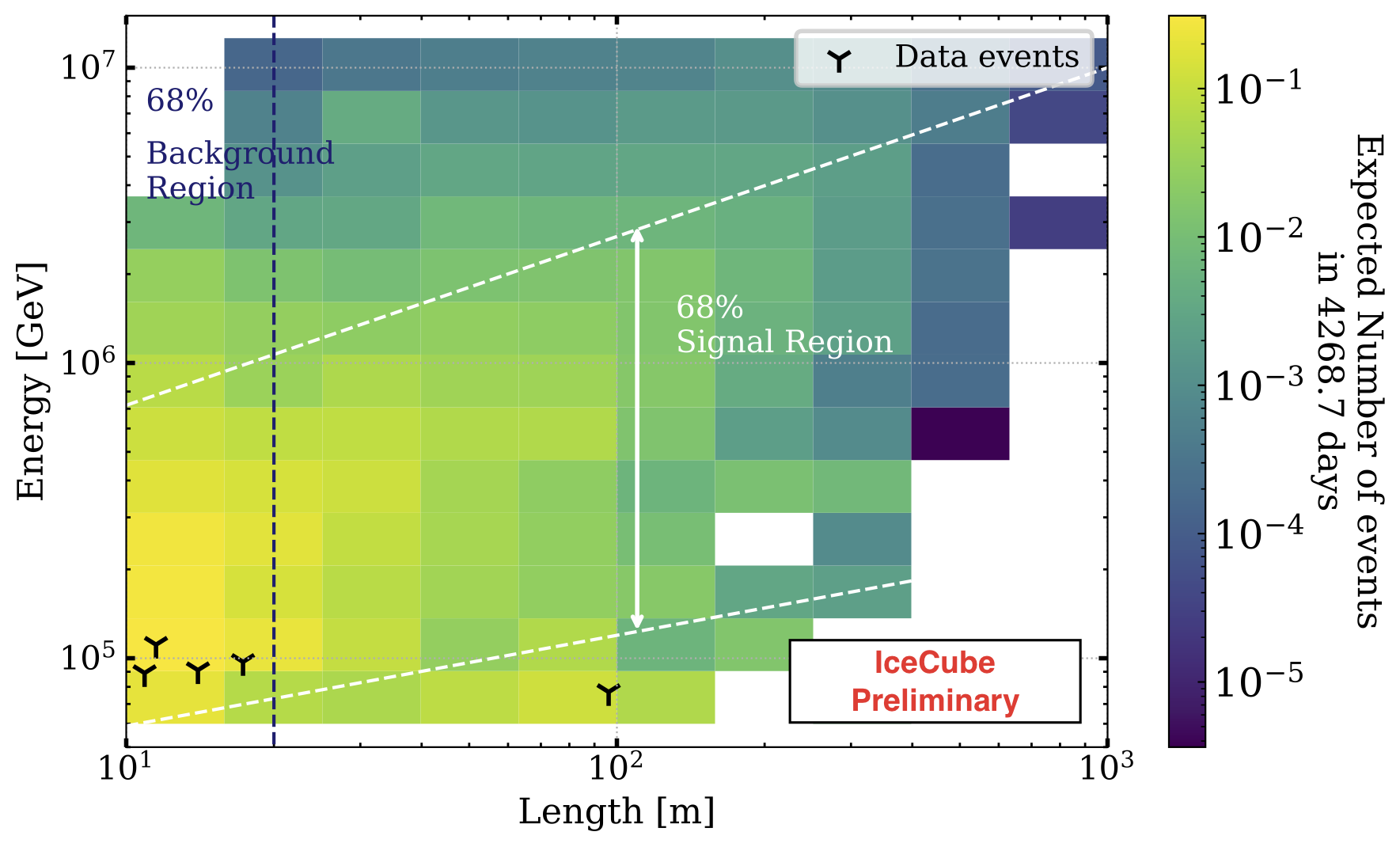}
  \caption{
  Expected number of double cascade candidate events according to the best-fit model as a function of reconstructed energy and tau length. The 5 observed data events are marked with black markers. The signal (white) and background (dark blue) dominated regions are marked with their respective percentiles.}
  \label{fig:datamcdoublecascade}
\end{SCfigure}

This analysis found five double cascade candidate events that are shown in Figure \ref{fig:datamcdoublecascade}, together with the best-fit Monte Carlo distribution. The plot displays both the data and the probability density function of the reconstructed energy and length. The expected background decreases with length and the vertical line indicates the region that contains 68\% of the background. Double cascade signals from $\nu_\tau$ CC interactions exhibit a strong correlation between reconstructed energy and length, with 68\% of the events falling between the diagonal lines. Four of the five events have reconstructed lengths below 17 m and fall in the 68\% background region. One event is reconstructed with a length of 96 m, but a reconstructed energy of 77 TeV. The combination of an arbitrary long length with a low reconstructed energy is typical for misclassified muons. The "Double Double" event reported in Ref. \cite{julianahese7.5} falls within the 68\% region for both signal and background, with a reconstructed tau decay length of $l_\tau = 17.3$ m and a total energy of $E_{\rm tot} = 97$ TeV.

Following three iterations of this analysis using 6, 7.5, and now 12 years of data, further progress will require significant changes to the analysis framework. While the current methodology has yielded meaningful results, it faces limitations. In particular, it is limited by the sparse Monte Carlo statistics for the double cascade sample. Additionally, the double cascade identification and reconstructed observables have shown limited robustness under changes to reconstruction algorithms and updates to the ice model. The next section outlines several improvements designed to overcome these challenges and to enhance both the sensitivity and stability of the analysis moving forward.

\section{Improvements \& Outlook}

The observables for the double cascade sample in this analysis were the reconstructed energy and length, but there are more parameters that could help distinguish the double cascade signal from the background. 

\begin{SCfigure}[1.2][h!]
  \centering
  \includegraphics[width=0.6\linewidth]{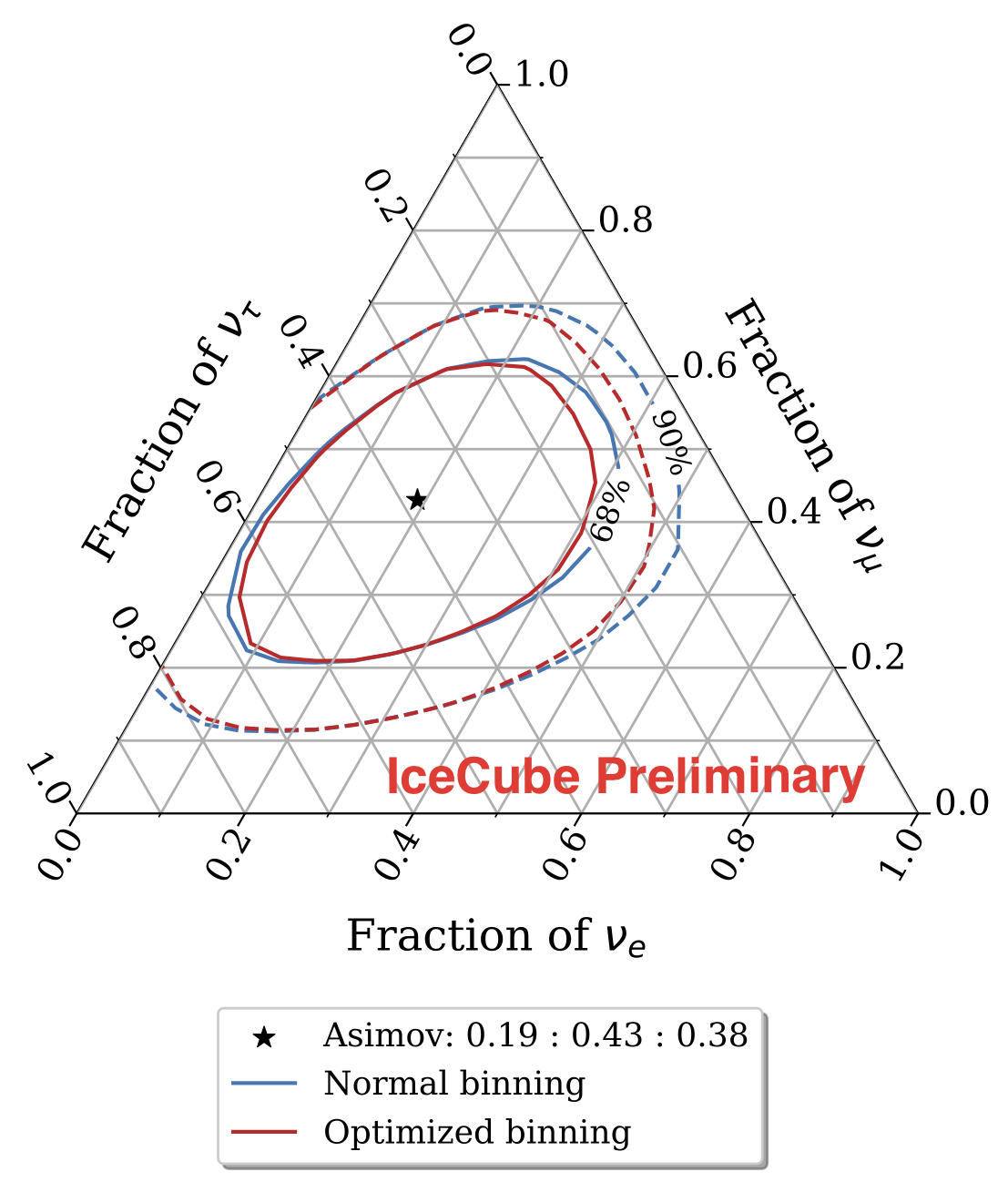}
  \caption{Projected sensitivity using normal (blue) and optimized (red) analysis approach. The solid (dashed) lines depict the corresponding 68\% (90\%) constraints derived using Wilks’ theorem.}
  \label{fig:sumstat}
\end{SCfigure}

Additional observables such as the energy asymmetry, the energy confinement and the zenith angle were not included because it would increase the dimensionality of the analysis histogram. Limited Monte Carlo statistics hinder an accurate estimation of the expected number of events, reducing the overall robustness of the analysis. This problem can be solved by employing an optimized summary statistic as described in Refs. \cite{infernosumstat,oliver}. This method trains a neural network to optimize a summary statistic, using the variance of the signal parameters as the loss function. This allows for the inclusion of more observables, while keeping enough Monte Carlo statistics per bin. The projected sensitivity using normal and optimized binning is shown in Figure \ref{fig:sumstat}. The normal binning entails 130 linear bins in $E_{\rm tot}$ and $l_\tau$, while the optimized binning was obtained by training on the model on $E_{\rm tot},l_\tau,\theta,\phi$ and dividing the summary statistic in 80 bins. The contours were obtained using Wilks' theorem and show improved constrains using less bins with the optimized summary statistic.

\begin{figure}[h!]
    \centering
    \begin{subfigure}[t]{0.49\linewidth}
        \centering
        \includegraphics[width=\linewidth]{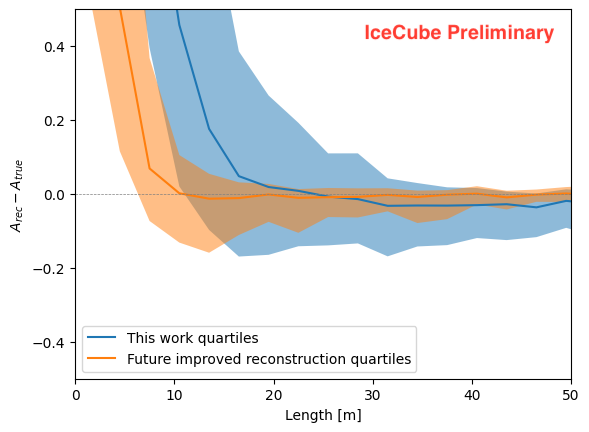}
        \caption{}
        \label{fig:recol}
    \end{subfigure}
    \hfill
    \begin{subfigure}[t]{0.49\linewidth}
        \centering
        \includegraphics[width=\linewidth]{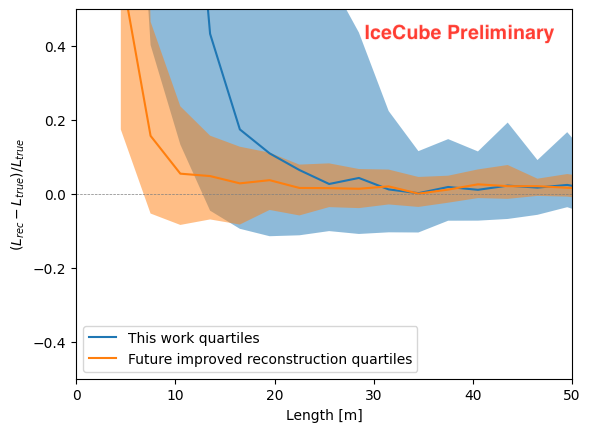}
        \caption{}
        \label{fig:recol_evtgen}
    \end{subfigure}
    \caption{
    The median and quartiles of the reconstructed energy asymmetry $A_{\rm rec}$ (left) and length $L_{\rm rec}$ (right) resolution for double cascades in the HESE sample.}
    \label{fig:recol_comparison}
\end{figure}

The reconstructed length and energy asymmetry are the most effective parameters to distinguish single cascades from double cascades. Recent improvements in the reconstruction algorithm, specifically in optimizing the starting point of the fit, have led to improved reconstruction performance. The median and quartiles of the reconstructed energy asymmetry $A_{\rm rec}$ and length $L_{\rm rec}$ resolution are shown in Figure \ref{fig:recol_comparison}. Both quantities show significant improvements for small tau lengths, which enhances the discrimination between double and single cascades.

Further improvements are anticipated by replacing the current particle identification method with a boosted decision tree, as demonstrated for the cascade sample in Ref. \cite{zheyang}. This will be followed by combining the HESE sample with the northern track sample \cite{abbasi2022improved} and the cascade sample \cite{cascade} to perform a combined measurement of the flavor composition. Additional sensitivity to the flavor composition is expected from cascades at lower energies, as shown by Medium Energy Starting Event Sample \cite{mese2year, meseicrc2025}.

The IceCube Upgrade, scheduled for deployment during the 2025–2026 polar season, will introduce a series of new calibration devices designed to improve our understanding and description of the optical properties of the ice \cite{upgrade}. In turn, this will enhance the reconstruction of tau neutrino events, which are particularly sensitive to the modeling of light propagation in the detector.

The Upgrade instrumentation will also include a densely instrumented volume that facilitates dedicated flasher studies designed to mimic double cascade signatures \cite{usner2018search}, providing a controlled environment for refining reconstruction techniques. Previous studies using the existing IceCube optical modules revealed anisotropic scattering in the ice \cite{anisotropy}, a feature now incorporated into current ice models.

With these advancements, future data — along with existing datasets — can be analyzed using improved reconstruction methods and updated calibrations, offering the potential for significantly enhanced sensitivity to tau neutrinos and the flavor composition.

\bibliographystyle{ICRC}
\bibliography{references}

%

\clearpage

\section*{Full Author List: IceCube Collaboration}

\scriptsize
\noindent
R. Abbasi$^{16}$,
M. Ackermann$^{63}$,
J. Adams$^{17}$,
S. K. Agarwalla$^{39,\: {\rm a}}$,
J. A. Aguilar$^{10}$,
M. Ahlers$^{21}$,
J.M. Alameddine$^{22}$,
S. Ali$^{35}$,
N. M. Amin$^{43}$,
K. Andeen$^{41}$
% ...
\\
\\
% ...
$^{10}$ Universit{\'e} Libre de Bruxelles, Science Faculty CP230, B-1050 Brussels, Belgium \\
% ...
$^{16}$ Department of Physics, Loyola University Chicago, Chicago, IL 60660, USA \\
$^{17}$ Dept. of Physics and Astronomy, University of Canterbury, Private Bag 4800, Christchurch, New Zealand \\
% ...
$^{21}$ Niels Bohr Institute, University of Copenhagen, DK-2100 Copenhagen, Denmark \\
$^{22}$ Dept. of Physics, TU Dortmund University, D-44221 Dortmund, Germany \\
% ...
$^{35}$ Dept. of Physics and Astronomy, University of Kansas, Lawrence, KS 66045, USA \\
% ...
$^{39}$ Dept. of Physics and Wisconsin IceCube Particle Astrophysics Center, University of Wisconsin{\textemdash}Madison, Madison, WI 53706, USA \\
% ...
$^{41}$ Department of Physics, Marquette University, Milwaukee, WI 53201, USA \\
% ...
$^{43}$ Bartol Research Institute and Dept. of Physics and Astronomy, University of Delaware, Newark, DE 19716, USA \\
% ...
$^{63}$ Deutsches Elektronen-Synchrotron DESY, Platanenallee 6, D-15738 Zeuthen, Germany \\
$^{\rm a}$ also at Institute of Physics, Sachivalaya Marg, Sainik School Post, Bhubaneswar 751005, India \\
% ...

\subsection*{Acknowledgments}

\noindent
The authors gratefully acknowledge the support from the following agencies and institutions:
USA {\textendash} U.S. National Science Foundation-Office of Polar Programs,
U.S. National Science Foundation-Physics Division,
U.S. National Science Foundation-EPSCoR,
U.S. National Science Foundation-Office of Advanced Cyberinfrastructure,
Wisconsin Alumni Research Foundation,
Center for High Throughput Computing (CHTC) at the University of Wisconsin{\textendash}Madison,
Open Science Grid (OSG),
Partnership to Advance Throughput Computing (PATh),
Advanced Cyberinfrastructure Coordination Ecosystem: Services {\&} Support (ACCESS),
Frontera and Ranch computing project at the Texas Advanced Computing Center,
U.S. Department of Energy-National Energy Research Scientific Computing Center,
Particle astrophysics research computing center at the University of Maryland,
Institute for Cyber-Enabled Research at Michigan State University,
Astroparticle physics computational facility at Marquette University,
NVIDIA Corporation,
and Google Cloud Platform;
Belgium {\textendash} Funds for Scientific Research (FRS-FNRS and FWO),
FWO Odysseus and Big Science programmes,
and Belgian Federal Science Policy Office (Belspo);
Germany {\textendash} Bundesministerium f{\"u}r Forschung, Technologie und Raumfahrt (BMFTR),
Deutsche Forschungsgemeinschaft (DFG),
Helmholtz Alliance for Astroparticle Physics (HAP),
Initiative and Networking Fund of the Helmholtz Association,
Deutsches Elektronen Synchrotron (DESY),
and High Performance Computing cluster of the RWTH Aachen;
Sweden {\textendash} Swedish Research Council,
Swedish Polar Research Secretariat,
Swedish National Infrastructure for Computing (SNIC),
and Knut and Alice Wallenberg Foundation;
European Union {\textendash} EGI Advanced Computing for research;
Australia {\textendash} Australian Research Council;
Canada {\textendash} Natural Sciences and Engineering Research Council of Canada,
Calcul Qu{\'e}bec, Compute Ontario, Canada Foundation for Innovation, WestGrid, and Digital Research Alliance of Canada;
Denmark {\textendash} Villum Fonden, Carlsberg Foundation, and European Commission;
New Zealand {\textendash} Marsden Fund;
Japan {\textendash} Japan Society for Promotion of Science (JSPS)
and Institute for Global Prominent Research (IGPR) of Chiba University;
Korea {\textendash} National Research Foundation of Korea (NRF);
Switzerland {\textendash} Swiss National Science Foundation (SNSF).

\end{document}